# Petrus Peregrinus of Maricourt and the Medieval Magnetism

Amelia Carolina Sparavigna

1 – Department of Applied Science and Technology, Politecnico di Torino, Torino, Italy

**ABSTRACT.** Petrus Peregrinus of Maricourt, a 13th-century French scholar and engineer, wrote what we can consider as the first extant treatise on magnetism of Europe. This treatise is in the form of a letter, probably composed during the siege of Lucera in Italy, in 1269, where Peregrinus worked to fortify the camp and built engines for projecting stones and fireballs into the besieged town. Peregrinus' letter consists of two parts. The first is discussing the properties of magnets, describing also the methods for determining their north and south poles. The second part of the letter describes some instruments that utilize the properties of magnets, ending with the Peregrinus' art of making a wheel of perpetual motion. In this paper, we discuss the first part of the letter and the related medieval knowledge of magnetism.

**Introduction:** Petrus Peregrinus of Maricourt was a 13th-century French scholar and engineer, that conducted and reported several experiments on magnetism. His abilities as an experimenter were well-known in that period and highly praised by one of his contemporaries, the English philosopher and Franciscan friar Roger Bacon. Peregrinus wrote what we can consider as the first extant treatise on magnets of Europe. This treatise is in the form of a letter, and it is entitled "Epistola Petri Peregrini de Maricourt ad Sygerum de Foucaucourt, Militem, de Magnete", "Letter of Peter Peregrinus of Maricourt to Sygerus of Foucaucourt, Soldier, on the Magnet". In one of the surviving manuscript copies, it is told that the letter was composed during the siege of Lucera in Italy, dated 8 August 1269. Probably, Petrus Peregrinus was in the army of Charles I, duke of Anjou and king of Sicily, who was besieging Lucera in a "crusade" sanctioned by the pope.

Peregrinus' Letter on the magnet consists of two parts. The first treats the properties of the lodestone (magnetite), providing a description of the polarity of magnets and methods for determining their north and south poles. In the first part, Peregrinus describes also the effects of attraction and repulsion between poles. The second parts of the Letter describes instruments that utilize the properties of magnets, in particular the floating compass, and proposes a new pivoted compass in some detail. The Letter ends with the Peregrinus' art of making a wheel of perpetual motion.

As we will see in the following discussion, some observations about magnets were existing in the medieval cultural environment. However, Peregrinus was able organizing the whole into a text that formed the basis of the science of magnetism. The Letter is generally considered as one of the great works of medieval experimental research, and, the methods exposed in it as precursors of modern scientific methodology [1]. We can find the Letter in the text entitled "Petrus Peregrinus on the Magnet, A.D. 1269" [2], translated from Latin by Brother Arnold (Joseph Charles Mertens [3]), Principal of La Salle Institute in Troy. The Letter was introduced by a discussion of Brother Potamian (Michael Francis O Reilly [4]), professor of Physics at Manhattan College of New York.

**Magnetism in classic antiquity and middle ages:** In the classic antiquity and in the medieval period, we can find several descriptions of the attraction which lodestone manifests for iron. In his introduction to Peregrinus' Letter [2], Brother Potamian writes that Lucretius (99-55 BC) gave a poetical dissertation on the magnet in his "De Rerum Natura", Book VI. Lucretius recognized magnetic repulsion, magnetic induction, and, according to Potamian, "to some extent the magnetic field with its lines of force". The poet Claudian (365-408 AD) wrote a short idyll on the attractive virtue of the lodestone and its symbolism; Saint Augustine (354-430 AD), in his work "De Civitate Dei", wrote that a lodestone, held under a silver plate, draws after it a scrap of iron lying on the plate [2,5]. It is also interesting to note that the Augustinian Abbot Alexander of Neckam (1157-1217) was distinguishing between the properties of the two ends of the lodestone. In his "De Utensilibus", Neckam provides what is perhaps the earliest reference to the mariner's compass in the Western Europe. In the world, it was a Chinese encyclopedist author, Shen Kuo, who gave the first known account of suspended magnetic compasses, a hundred years earlier, in 1088 AD, in the book entitled "Meng Xi Bi Tan" (Dream Pool Essays) [6].

The Dominican friar and bishop Albertus Magnus (1193-1280), in his treatise "De Mineralibus", describes several kind of magnets and states some of the properties commonly attributed to them [2]. The minstrel Guyot de Provins, in a satirical poem written about 1208, refers to the directive quality of the lodestone and its use in navigation [2,7]. We find the magnetic compass also in the "Historia Orientalis" (1215-1220) by Cardinal Jacques de Vitry, in the "Tresor des Sciences" (1260) written in Paris by Brunetto Latini, poet and philosopher, in a treatise written by the 'Enlightened Doctor' Raymond Lully, and in the famous canzone "Al cor gentil rempaira sempre amore" (Love always has its home in the noble heart), composed by Guido Guinizelli, the poet-priest of Bologna [2]. In Ref.1 we find mentioned other scholars too. Bartholomaeus Anglicus (1220-1250) refers to the magnet in his encyclopedic treatise "De proprietatibus rerum" (On the properties of things). Henry Bate (1246-1317) included a substantial discussion of magnetism in his "Speculum divinorum et quorundam naturalium" (Mirror of divine things and of some natural ones).

**Magnets and diamonds:** Let us discuss for a while the reference to Guinizelli's poetry. In his canzone on love, the poet tells "Amore in gentil cor prende rivera per suo consimel loco com'adamas del ferro in la minera," that is "love has home in a gently noble heart, like, in the same manner, adamas has home in an iron mine." What is the "adamas"? Some commentators translate it as "diamond", others, probably more correctly, as "magnet" [8]. In fact, the word "adamas" is the medieval word for both lodestone and diamond. In the Guinizelli's canzone, when we consider "adamas" as "magnet", we have a clear example of the medieval similitude between "love" and "magnet" that was common in troubadour lyrics. For instance: "tira com azimans, la bela", that is, "the fair lady draws me toward her like a magnet", writes Bernart de Ventadorn [9]. The similitude is reinforced by a phonetic resemblance between the words for "magnet" and "love". As noted in [8], for medieval poets the true lover (amans) was like a magnet (azimans, adamas).

In the book of William Gilbert (1544-1603), English physicist and natural philosopher, on magnetism [10], we find several names for magnets from different countries. Gilbert writes that in English, the magnet is known as "lodestone" and "adamant stone" (William Shakespeare used "adamant" too, in the Midsummer Night's Dream: "You draw me, you hard-hearted adamant, but yet you draw not iron; for my heart is true as steel"). Adamant is another form of "adamas". In various forms (adamas, adamant, aimant, azimans, aymant, yman) and in many languages, we find the original ancient Greek "adamas", the "unconquered". Originally, the word was applied by the Greeks to the hardest of the metals with which they were acquainted, that is to say, to hard-tempered iron or steel. Due to its meaning, this word was subsequently applied to diamond for the same reason. In the writings of the middle ages, and even in Pliny the Elder, we find some confusion between the two uses of "adamas" to denote the lodestone as well as the diamond [10].

**Petrus Peregrinus and the perpetual motion:** As told in [2], of the early years of Peregrinus nothing is known. He studied probably at the University of Paris and graduated with the highest scholastic honors. His surname is coming from the village of Maricourt, in Picardy, whereas the appellation Peregrinus, or Pilgrim, is due to the fact that he visited the Holy Land. He was also known as 'Peter Adsiger', as we can find in a book of 1787, written by Tiberio Cavallo (1749-1809), entitled "Treatise on Magnetism" (London) [11].

In 1269, we find Peter Peregrinus in the engineering corps of the French army that was besieging Lucera, in Southern Italy. Peregrinus worked to fortify the camp and lay mines. He also worked to build engines for projecting stones and fireballs into Lucera. It seems that, during such warlike preoccupations, an idea occurred to Peregrinus: the idea was of devising a mechanism able of keeping the astronomical "sphere" of Archimedes in uniform rotation [2].

Of the "spheres" of Archimedes, wrote Cicero, the Roman philosopher and politician, in the first century BC. Cicero wrote of two spherical objects built by Archimedes, that Marcellus, the Roman consul who conquered Syracuse in 212 BC, brought to Rome [12]. One was a solid sphere on which were engraved or painted stars and constellations; the second sphere was much more ingenious and original. It was a planetarium, a mechanical device showing the motions of sun, moon, and planets as viewed from Earth. No physical trace of Archimedes' planetarium survives, but we can have some ideas about it. In 1900, a shipwreck found near the Greek island of Antikythera uncovered an exceptional object. Amidst the cargo of a ship dated from the first century BC, there was a small lump of wood and corroded gears of bronze, which revealed itself as an analog computer designed to predict astronomical positions and eclipses. The device is known as the 'Antikythera mechanism' [13,14]. Of course, we cannot attribute this mechanics to Archimedes, but we can imagine he could had built a similar device too, that the consul Marcellus brought to Rome. And in fact, recently, a model of Archimedes' sphere had been reconstructed by Michael Wright, who was a curator at the Science Museum in London and that spent many years studying the Antikythera mechanism. His globe, made from copper and brass displays the movements of the sun, moon and planets as they travel through the night sky [15].

Peregrinus, attracted by the mechanical problems connected with Archimedes' planetarium, was gradually led to consider the problem of perpetual motion. The result was that he described, "to his own evident satisfaction," [2] how a wheel might be driven round forever by the power of magnetic attraction. "Elated over his imaginary success," Peregrinus wrote to inform a friend at home. To allow his friend comprehending the mechanism of the motor and the functions of its parts, "he proceeds to set forth in a methodical manner all the properties of the lodestone, most of which he himself had discovered." [2]

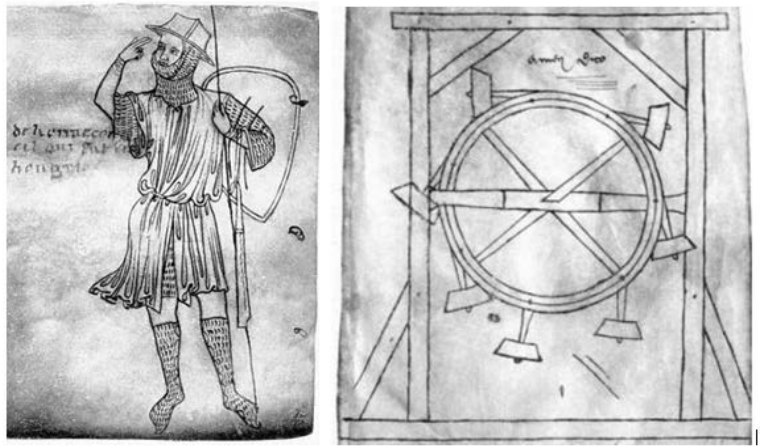

*Fig. 1. Two drawings from the notebook of Villard de Honnecourt, an artist from Picardy, contemporary of Peter Peregrinus. A drawing is showing how could appear a soldier at the time. On the right, we can see a wheel of perpetual motion as imagined by Villard de Honnecourt.*

Peter Peregrinus was not the only person from Picardy that studied the problem of perpetual motion. Another one was his contemporary Villard de Honnecourt. Villard is known to history only through a surviving notebook of 33 sheets of parchment containing about 250 drawings dating from the 1220s/1240s, which is now in the Bibliothèque Nationale, Paris (MS Fr 19093). The great variety of subjects (religious and secular figures, architectural plans and mechanical devices), makes it difficult to determine his profession. Since the discovery of his notebook, it is general opinion that Villard was an itinerant architect. Among the mechanical devices sketched by Villard, we see the perpetual-motion machine shown in the Figure 1. The problem of perpetual motions was of great appeal during the middle ages. This interest was probably stimulated by the books on mechanics coming from Arabic world, where we can find such wheels. The wheel sketched by Peregrinus is discussed in a very interesting article [16], which is also showing several layouts of it in different manuscripts and also wheels from Arabic manuscripts.

**Roger Bacon's opinion**: Peregrinus' Letter was the first landmark among the studies on magnetism, the next being William Gilbert's De Magnete, in 1600. The Letter was addressed to Sigerus de Foucaucourt, his "amicorum intimus," the dearest of friends. Another friend was Roger Bacon, who held Peregrinus in the very highest esteem, as shows by his following words: "There are but two perfect mathematicians," wrote the English monk, "John of London and Petrus de Maharne-Curia, a Picard" [2]. Bacon thus writes of Peregrinus [2]: "I know of only one person who deserves praise for his work in experimental philosophy, for he does not care for the discourses of men and their wordy warfare, but quietly and diligently pursues the works of wisdom. … he is a master of experiment. … he knows all natural science whether pertaining to medicine and alchemy, or to matters celestial and terrestrial. He has worked diligently in the smelting of ores as also in the working of minerals; he is thoroughly acquainted with all sorts of arms and implements used in military service and in hunting, besides which he is skilled in agriculture and in the measurement of lands. It is impossible to write a useful or correct treatise in experimental philosophy without mentioning this man's name". Other references and information on Petrus Peregrinus are reported in [17].

**Analysis of the Letter**: The analysis proposed in [2] shows that, according to the known manuscripts: 1) Peter Peregrinus was the first to assign a definite position to the poles of a lodestone and to provide a method for determining which is north and which south; 2) he proved that unlike poles attract each other, and that similar ones repel; 3) after experiments, he established every fragment of a lodestone, however small, has two poles and then it is a complete magnet; 4) he recognized that a pole of a magnet may neutralize a weaker one of the same name, and even reverse its polarity; 5) he was the first to describe the use of a pivot for a magnetized needle and surround it with a graduated circle, creating, in such a manner, a model for the modern magnetic compass; 6) he determined the position of an object by its magnetic bearing as done in modern compass surveying; and, at the end of the letter, 7) he described his perpetual motion machine, based on the idea of a magnetic motor, a clever and new idea for a thirteenth century engineer [2].

The copies of Peregrinus' Letter for nearly three centuries, remained unnoticed among the libraries of Europe, until William Gilbert, who makes frequent mention of it, published his "De Magnete" in 1600. After, a Jesuit writer, Niccolò Cabeo, refers to it in his "Philosophia Magnetica", 1629. And Athanasius Kirches quotes from the Letter, in his "De Arte Magnetica", 1641. Kircher also constructed a magnetic clock, the mechanism of which is described in his book.

**In the first part of the Letter:** After an introduction, where Peregrinus writes that he wants to explain to his friend the hidden virtue of the lodestone in a simple style, he poses the "qualifications of the experimenter". "Whoever wishes to experiment, should be acquainted with the nature of things, and should not be ignorant of the motion of the celestial bodies. He must also be skilful in manipulation in order that, by means of this stone, he may produce these marvelous effects… Besides, in such occult experimentation, great skill is required, for very frequently without it the desired result cannot be obtained, because there are many things in the domain of reason which demand this manual dexterity" [2].

After the experimenter has found a good lodestone, he can find and distinguish its poles. "I wish to inform you that this stone bears in itself the likeness of the heavens, as I will now clearly demonstrate." That is, we have in the heavens "two points more important than all others, because on them, as on pivots, the celestial sphere revolves:" these points are the Arctic or north pole and the Antarctic or south pole. The lodestone has two points which are respectively the north pole and the south pole. "If you are very careful, you can discover these two points in a general way. One method for doing so is the following: with an instrument with which crystals and other stones are rounded, let a lodestone be made into a globe and then polished. A needle or an elongated piece of iron is then placed on top of the lodestone and a line is drawn in the direction of the needle or iron, thus dividing the stone into two equal parts. The needle is next placed on another part of the stone and a second median line drawn. If desired, this operation may be performed on many different parts, and undoubtedly all these lines will meet in two points just as all meridian or azimuth circles meet in the two opposite poles of the globe. One of these is the north pole, the other the south pole." [2] In fact, Peter Peregrinus is telling that it is possible to create a globe and, on it, finding the poles by drawing on it a set of meridians, which are following the lines of the magnetic field, detected by means of a needle (see Figure 2).

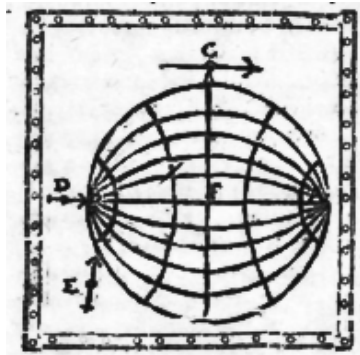

*Fig. 2. A spherical magnet with poles and meridians, as illustrated in the "Tractatus, sive Physiologia nova de magnete, magneticisque corporibus & magno magnete tellure" by William Gilbert, published 1633 by Lochmans.*

In the Figure 2, the spherical magnet looks like a "terrella", Latin of "little earth", a small magnetised model representing the Earth. Terrella is usually thought to have been invented by William Gilbert, but based on an idea of Peter Peregrinus.

Peter Peregrinus is describing another method for determining the poles. "Note the place on the above-mentioned spherical lodestone where the point of the needle clings most frequently and most strongly; for this will be one of the poles as discovered by the previous method. In order to determine this point exactly, break off a small piece of the needle or iron so as to obtain a fragment about the length of two fingernails; then put it on the spot which was found to be the pole by the former operation (see Figure 2). If the fragment stands perpendicular to the stone, then that is, unquestionably, the pole sought; if not, then move the iron fragment about until it becomes so; mark this point carefully; on the opposite end another point may be found in a similar manner. If all this has been done rightly, and if the stone is homogeneous throughout and a choice specimen, these two points will be diametrically opposite, like the poles of a sphere" [2].

**North and South Poles:** After we have found the poles, we have to determine which is north and which south. We can proceed in the following manner, according to Peregrinus. He is proposing to use the celestial pole as a reference. Let us take a wooden vessel, made like a dish, and place in it the stone in such a way that the two poles will be equidistant from the edge of the vessel. Then, let us place the dish in another and larger vessel full of water, so that "the stone in the first-mentioned dish may be like a sailor in a boat". The second vessel should be of considerable size, in order that the lodestone may not be impeded by contact of one vessel against the sides of the other. "When the stone has been thus placed, it will turn the dish round until the north pole lies in the direction of the north pole of the heavens, and the south pole of the stone points to the south pole of the heavens". And then "since the north and south parts of the heavens are known, these same points will then be easily recognized in the stone because each part of the lodestone will turn to the corresponding one of the heavens," Peregrinus explains.

**How lodestones attract each other:** After we discovered the north and the south pole in the lodestone, we have to mark them both carefully. If we want to see how one lodestone attracts another, then, with two lodestones selected and prepared as previously told, we can proceed as follows. "Place one in its dish that it may float about as a sailor in a skiff, and let its poles which have already been determined be equidistant from the horizon, i.e., from the edge of the vessel. Taking the other stone in your hand, approach its north pole to the south pole of the lodestone floating in the vessel; the latter will follow the stone in your hand as if longing to cling to it. If, conversely, you bring the south end of the lodestone in your hand toward the north end of the floating lodestone, the same phenomenon will occur; namely, the floating lodestone will follow the one in your hand. Know then that this is the law: the north pole of one lodestone attracts the south pole of another, while the south pole attracts the north. Should you proceed otherwise and bring the north pole of one near the north pole of another, the one you hold in your hand will seem to put the floating one to flight. If the south pole of one is brought near the south pole of another, the same will happen. This is because the north pole of one seeks the south pole of the other, and therefore repels the north pole". [2]

After the discussion of this experiment, Peregrinus continues remarking that it "is well known to all who have made the experiment, that when an elongated piece of iron has touched a lodestone and is then fastened to a light block of wood or to a straw and made float on water, one end will turn to the star which has been called the Sailor's star because it is near the pole; the truth is, however, that it does not point to the star but to the pole itself". Peregrinus is also telling an important fact, that every fragment of a lodestone has two poles and then it is a complete magnet. "Take a lodestone which you may call AD, in which A is the north pole and D the south; cut this stone into two parts, so that you may have two distinct stones; place the stone having the pole A so that it may float on water and you will observe that A turns towards the north as before; the breaking did not destroy the properties of the parts of the stone, since it is homogeneous; hence it follows that the part of the stone at the point of fracture, which may be marked B, must be a south pole; this broken part of which we are now speaking may be called AB. The other, which contains D, should then be placed so as to float on water, when you will see D point towards the south because it is a south pole; but the other end at the point of fracture, lettered C, will be a north pole; this stone may now be named CD. If we consider the first stone as the active agent, then the second, or CD, will be the passive subject. You will also notice that the ends of the two stones which before their separation were together, after breaking will become one a north pole and the other a south pole. If now these same broken portions are brought near each other, one will attract the other, so that they will again be joined at the points B and C, where the fracture occurred. Thus, by natural instinct, one single stone will be formed as before." [2].

**The natural virtue of magnets:** In the part of the Letter discussing the natural virtue of magnets, we can find an experimental device, which Peregrinus is proposing for having a magnetic clock. To Peregrinus, "it is clear that the poles of the lodestone derive their virtue from the poles of the heavens. As regards the other parts of the stone, the right conclusion is that they obtain their virtue from the other parts of the heavens. … You may test this in the following manner: A round lodestone on which the poles are marked is placed on two sharp styles as pivots having one pivot under each pole so that the lodestone may easily revolve on these pivots. Having done this, make sure that it is equally balanced and that it turns smoothly on the pivots. … Then place the stone with its axis in the meridian, the poles resting on the pivots. Let it be moved after the manner of bracelets (the bracelets are the circles, "armillae", of armillary spheres) so that the elevation and depression of the poles may equal the elevation and depressions of the poles of the heavens of the place in which you are experimenting. If now the stone be moved according to the motion of the heavens, you will be delighted in having discovered such a wonderful secret; … With such an instrument you will need no timepiece, for by it you can know the ascendant at any hour you please, as well as all other dispositions of the heavens which are sought for by astrologers" [2].

About this experiment, William Gilbert tells in his book [10]: "I omit what Peter Peregrinus constantly affirms, that a terrella suspended above its poles on a meridian moves circularly, making an entire revolution in 24 hours: which, however, it has not happened to ourselves as yet to see …". A comment in [10] tells us that, besides Gilbert, Galileo too discussed this experiment in the third of his Dialogues, the book which presents a series of discussions among two philosophers and a layman:

Salviati, who presents some of Galileo's views directly, Sagredo and Simplicio, a follower of Ptolemy and Aristotle. About Peregrinus' experiment, Salviati tells "I will speak to one particular, to which I could have wished, that Gilbert had not lent an ear; I mean that of admitting, that in case a little Sphere of Loadstone might be exactly librated, it would revolve in itself; because there is no reason why it should do so; For if the whole Terrestrial Globe hath a natural faculty of revolving about its own centre in twenty four hours, and that all its parts ought to have the same, I mean, that faculty of turning round together with their whole, about its centre in twenty four hours; they already have the same in effect, whilst that, being upon the Earth, they turn round along with it: And the assigning them a revolution about their particular centres, would be to ascribe unto them a second motion much different from the first; for so they would have two, namely, the revolving in twenty four hours about the centre of their whole; and the turning about their own: now this second is arbitrary, nor is there any reason for the introducing of it" [18].

With the discussion of the pivoted sphere made by Galileo, let us conclude this discussion on the medieval magnetism as we find in the Peregrinus' Letter. However, let us stress that the attraction the Peregrinus had for pivoted magnets, forced him to imagine new devices. In the second part of the Letter, he discussed three devices: one is an instrument for measuring the azimuth of sun, moon and stars on the horizon, the second a pivoted compass and the third a wheel of perpetual motion. The use of a pivoted compass to determine the azimuth of the sun is clearly shown by the Figure 3, which is obtained adapting images from [2]. The devices described by the Peregrinus will be the subject of a future paper, on pivoted mechanisms of the Middle Ages.

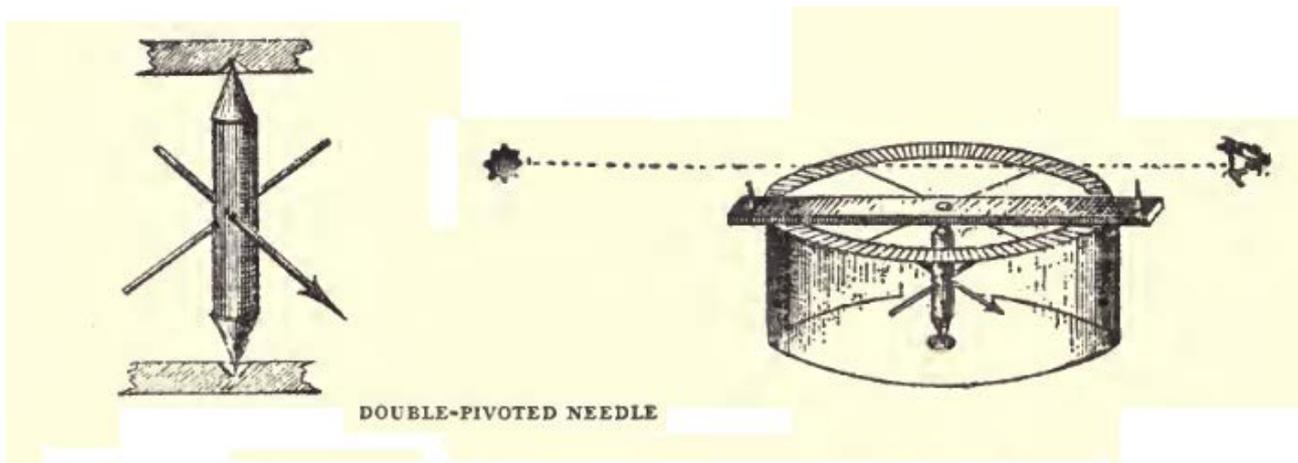

*Fig. 3. The pivoted magnet used for measuring the azimuth of stars.*